\documentclass[
preprint,
 amsmath,amssymb,
 aps,
prb,
]{revtex4-2}

\usepackage{booktabs,multirow,tabularx}

\usepackage[margin=1in]{geometry}
\usepackage{amsmath}
\usepackage{amssymb}
\usepackage{amsthm}
\usepackage{amsfonts}
\usepackage{listings}
\usepackage{enumerate}
\usepackage{latexsym}
\usepackage{psfrag}
\usepackage{bm}
\usepackage{graphicx}
\usepackage[caption=false]{subfig}
\usepackage{blkarray}
\usepackage{array}
\usepackage{color}
\usepackage[normalem]{ulem}
\usepackage[colorlinks,linkcolor=blue,citecolor=blue,urlcolor=blue]{hyperref}
\usepackage{dsfont}
\usepackage{ulem}
\usepackage{bbold}
\usepackage{siunitx}
\usepackage{physics}
\usepackage{comment}
\usepackage{mathtools}
\usepackage{sidecap}
\usepackage{stmaryrd}
\usepackage{inputenc }

\newcommand{\beq}{\begin{equation}}
	\newcommand{\eeq}{\end{equation}}
\newcommand{\beqarray}{\begin{eqnarray}}
	\newcommand{\eeqarray}{\end{eqnarray}}

\usepackage{titlesec}
\newcommand{\beginsupplement}{%
        \setcounter{secnumdepth}{3}
        \setcounter{tocdepth}{3} 
        \setcounter{equation}{0}
        \renewcommand{\theequation}{S\arabic{equation}}%
        \setcounter{figure}{0}
        \renewcommand{\thefigure}{S\arabic{figure}}%
        \renewcommand{\thesubsection}{S\arabic{subsection}}
     }

\begin{document}

\title{Non-Hermitian topological devices with Chern insulators}

\author{Kyrylo Ochkan}
\affiliation{Leibniz Institute for Solid State and Materials Research,
IFW Dresden, Helmholtzstrasse 20, 01069 Dresden, Germany}
\affiliation{W\"{u}rzburg-Dresden Cluster of Excellence ct.qmat, 01062 Dresden, Germany}

\author{Michael Wissmann}
\affiliation{Leibniz Institute for Solid State and Materials Research,
IFW Dresden, Helmholtzstrasse 20, 01069 Dresden, Germany}
\affiliation{Université Grenoble Alpes, CNRS, CEA, Spintec, F-38000 Grenoble, France}

\author{Louis Veyrat}
\affiliation{Laboratoire National des Champs Magnetiques Intenses LNCMI, CNRS-INSA-UJF-UPS, UPR3228, F-31400 Toulouse, France}

\author{Lixuan Tai}
\affiliation{RIKEN Center for Emergent Matter Science (CEMS), Wako, Saitama 351-0198, Japan}

\author{Minoru Kawamura}
\affiliation{RIKEN Center for Emergent Matter Science (CEMS), Wako, Saitama 351-0198, Japan}

\author{Yoshinori Tokura}
\affiliation{RIKEN Center for Emergent Matter Science (CEMS), Wako, Saitama 351-0198, Japan}


\author{Viktor K\"{o}nye}
\affiliation{Leibniz Institute for Solid State and Materials Research,
IFW Dresden, Helmholtzstrasse 20, 01069 Dresden, Germany}
\affiliation{W\"{u}rzburg-Dresden Cluster of Excellence ct.qmat, 01062 Dresden, Germany}

\author{Bernd Büchner}
\affiliation{Leibniz Institute for Solid State and Materials Research,
IFW Dresden, Helmholtzstrasse 20, 01069 Dresden, Germany}
\affiliation{Department of Physics, TU Dresden, D-01062 Dresden, Germany}

\author{Jeroen van den Brink}
\affiliation{Leibniz Institute for Solid State and Materials Research,
IFW Dresden, Helmholtzstrasse 20, 01069 Dresden, Germany}
\affiliation{Department of Physics, TU Dresden, D-01062 Dresden, Germany}

\author{Ion Cosma Fulga}
\affiliation{Leibniz Institute for Solid State and Materials Research,
IFW Dresden, Helmholtzstrasse 20, 01069 Dresden, Germany}
\affiliation{W\"{u}rzburg-Dresden Cluster of Excellence ct.qmat, 01062 Dresden, Germany}

\author{Joseph Dufouleur}
\affiliation{Leibniz Institute for Solid State and Materials Research,
IFW Dresden, Helmholtzstrasse 20, 01069 Dresden, Germany}
\affiliation{W\"{u}rzburg-Dresden Cluster of Excellence ct.qmat, 01062 Dresden, Germany}

\author{Romain Giraud}
\email{romain.giraud@cea.fr}
\affiliation{Leibniz Institute for Solid State and Materials Research,
IFW Dresden, Helmholtzstrasse 20, 01069 Dresden, Germany}
\affiliation{Université Grenoble Alpes, CNRS, CEA, Spintec, F-38000 Grenoble, France}


\begin{abstract}

{\bf Multi-terminal topological devices are a new generation of electronic devices with quantized properties robust against imperfections. In magnetic topological insulators, dissipationless edge states give functional devices in zero magnetic field, of interest for quantum metrology (resistance standard) or topological electronics (Chern networks).  
Here we show that a new generation of simple quantum circuits (disk, ring) with non-Hermitian topology, based on the interconnection of 1D Chern states in the quantum anomalous Hall regime, can have a much stronger quantization of their invariant than that of the Chern invariant itself, when measured in a non-metrology grade setup - that is, in industry-relevant conditions. Remarkably, the chirality-related topological skin effect is realized without the need of a magnetic field or an electrical gate, with a record degree of localization for a quantum Hall device. This new type of topological quantum devices based on magnets, with an exponential response that can be switched at small magnetic fields ($B\approx200$~mT), can operate at liquid-Helium temperature with a good quantization and have some potential as cryogenic sensors for applications in high-precision impedance or magnetic field measurements.}  

\end{abstract}

\maketitle

\newpage


\section{\label{Intro}Introduction}

{\bf Quantum devices based on Chern insulators}

In Chern insulators, conductance occurs via chiral edge channels and charge transport is ballistic. Due to their topological nature, the transverse (Hall) resistance is quantized, as defined by the Chern invariant $C$ (integer) and the von Klitzing constant $R_K=h/e^2$ only. In particular, this topological state can be achieved without applying a magnetic field, the time-reversal symmetry being broken by exchange fields or strong electronic interactions, which realizes the quantum anomalous Hall (QAH) effect \cite{Haldane1988,Yu2010,Chang2023}. Remarkable advances in materials' research on magnetic topological insulators \cite{Chang2013,Checkelsky2014,Mogi2015,Bestwick2015,Winnerlein2017} have enabled the improvement of the QAH response in macroscopic Hall-bar devices, with a precision at the metrology-grade level \cite{Bestwick2015,Goetz2018,Fox2018,Okazaki2020,Okazaki2022,Rodenbach2022,Patel2024}, now considered by metrology institutes as a quantum resistance standard in zero magnetic field compatible with other quantum technologies based on superconductors \cite{Poirier2025,Rodenbach2025}. Another application foreseen for topological electronics is based on Chern networks \cite{Gilbert2025}, considering computing based on complex multi-channel geometries of inhomogeneous, locally-controlled Chern insulators. Yet, the performance of quantum devices based on QAH materials remains strongly limited by temperature (below 1K), and the high-accuracy measurement of the Chern invariant requires advanced experimental setups with low-temperature filters and, ultimately, the use of ultra-stable cryogenic current comparators at metrology institutes \cite{Goetz2018,Fox2018,Okazaki2020,Okazaki2022,Rodenbach2022,Patel2024,Rodenbach2025}.


In this work, we show that a new generation of QAH-based topological devices built from the interconnection of Chern states in a chain geometry, as implemented in multi-terminal devices with simple shapes (disk, ring), have an intrinsic non-Hermitian topology with more robust quantized properties at elevated temperatures, tunable with small magnetic fields. Remarkably, in a simple lock-in amplifiers measurement setup, without low-temperature filters, these topological devices outperform other multi-terminal devices based on the intrinsic properties of Chern states only, with an accuracy of the non-Hermitian topological invariant that largely exceeds that of the Chern invariant. 
Moreover, the topological response of non-Hermitian devices, the skin effect, shows a record amplitude, with a gain of two orders of magnitude with respect to other quantum Hall platforms \cite{Ochkan2024,Oezer2024}.  Such quantum devices based on magnetic topological insulators thus have some great potential for applications as highly-sensitive cryogenic sensors, such as high-impedance sensors thanks to their exponential sensitivity to boundary conditions \cite{Budich2020,Konye2024}, or magnetic-field sensors based on their tunable topological response with small magnetic fields. 
Novel QAH-based non-Hermitian sensors are functional in zero magnetic field and can operate in small fields at liquid-helium temperature, whereas topological devices solely based on the quantization of Chern states work at sub-Kelvin temperatures only.

\section{\label{Results}Results}

\begin{figure}[!h]
\includegraphics[width=\columnwidth]{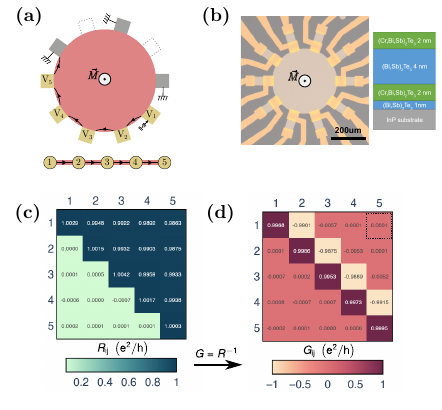}
\centering
\caption{
\textbf{Disk-shape quantum anomalous Hall device and conductance matrix.} 
\textbf{(a)} 5-site configuration where the grounded arms (grey contacts) are used to disconnect the first and the last site of the chain, thus realizing the open-chain boundary condition. The effective corresponding HN chain is shown schematically below. Active arms are labeled using currents and voltages, $I_j$ and $V_j$. For inactive arms in dashed lines, both contacts are floating.
\textbf{(b)} Left, Optical microscopy picture of the QAH disk device, with ten pairs of ohmic contacts; Right, Schematic of the heterostructure realizing the Chern insulator state without the need to apply an electrical gate voltage, as detailed in ref.~\cite{Okazaki2022}.
\textbf{(c)} Measured resistance matrix $[R_{i,j}]$ at $B = 0$~T and $T=30$~mK. 
\textbf{(d)} Calculated conductance matrix $[G_{i,j}]$ constructed by inverting the resistance matrix. 
}
\label{fig1}
\end{figure}


\textbf{Non-Hermitian topology of quantum anomalous Hall multi-terminal devices}

Chirality is a key ingredient to realize non-Hermitian quantum systems, as recently evidenced in a multi-terminal devices 
in the quantum Hall regime of 2D conductors \cite{Ochkan2024,Oezer2024}. Chiral edge states in the quantum Hall regime are perfectly transmitted between successive ohmic contacts, so that a simple device, with a disk or a ring geometry, realizes a 1D chain of dissipationless Chern states interconnected between dissipative contacts. As shown in Fig.~\ref{fig1}a), inner ohmic contacts are equivalent to sites between which chiral 1D states propagate. For a perfect transmission, this situation is equivalent to the non-Hermitian Hatano-Nelson model \cite{Hatano1996}, which results in the non-trivial topology of the conductance tensor of multi-terminal quantum Hall devices \cite{Ochkan2024}. Its specific signature, known as the topological skin effect, is the exponential localization of quantum states at one side of an open chain. Since magnetic topological insulators heterostructures realize quantum Hall edge states without the need of a magnetic field, QAH-based multi-terminal devices offer
new ways to implement non-Hermitian topology in cryogenic quantum electronics, with the additional possibility to switch chirality by applying small magnetic fields. To investigate their properties, such devices were built from a Cr-based diluted magnetic semiconductor heterostructure \cite{Okazaki2022}, patterned into a ring or a disk 
connected to pairs of ohmic contacts (Fig.~\ref{fig1}b). For each pair, the inner contact is used either for the current injection into chiral edge states or as a voltage probe along the chain, and some contacts can be used to open the chain, if connected to the ground. Outer contacts are either floating (open-boundary condition) or all connected to the ground (periodic-boundary condition).  
The size of the equivalent Hatano-Nelson chain is fixed by the number of contacts used to determine the conductance matrix. For instance, Fig.~\ref{fig1}a sketches a 5-terminal setup, equivalent to a 5-site quantum chain. By considering all possible current-injection paths, it is possible to measure the full resistance matrix $[R_{i,j}]$, as shown in Fig.~\ref{fig1}c) at $B=0$ for a remnant magnetization saturated perpendicular to the sample plane (QAH state with a given chirality), which is then converted into a conductance matrix $[G_{i,j}]$ by numerical calculations (Fig.~\ref{fig1}d).

\begin{figure}[!t!h]
\includegraphics[width=\columnwidth]{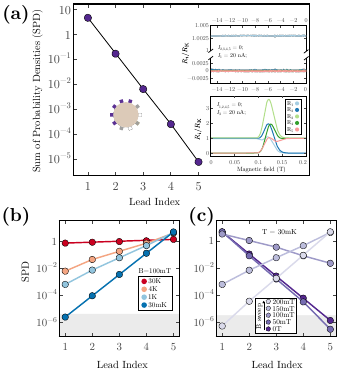}
\centering
\caption{
\textbf{Topological skin effect in a QAH disk.} \\
\textbf{(a)} Exponential localization of the sum of propability densities (SPD), the signature of the non-Hermitian topological skin effect, as measured at $T=30$~mK in zero magnetic field (after saturation of the magnetization in large negative fields) for the 5-site quantum chain; Upper inset: High-field dependence of the device resistances $R_i=V_i/I$, normalized to the von Klitzing constant $R_K$, measured in the current-injection scheme shown in Fig.~\ref{fig1}a), corresponding to the first raw in Fig.~\ref{fig1}c) and showing the stability of dissipationless edge states up to 15T; Lower inset: Low-field switching of the resistances measured with the current injected into the central contact $\#3$, showing the chirality switch at the magnetization reversal.  
\textbf{(b)} Temperature dependence of the SPD measured for the opposite direction of the saturated magnetization (after saturation of the magnetization in large positive fields), as measured under a small +100mT applied field so as to avoid the magnetization reversal at lower fields above 4K. This directly reveals the influence of the reversed chirality on the skin effect as well as its robustness at higher temperatures. 
\textbf{(c)} Rapid evolution of the skin effect upon magnetization reversal at small fields, after saturation of the magnetization in large negative fields. The skin effect is fully reversed in a narrow 20~mT field range, with a maximum slope $2.10^{-1}$~mT$^{-1}$, and it shows a record degree of localization for a 5-site quantum chain, with an improvement of more than two orders of magnitude with respect to AlGaAs-based devices.
}
\label{fig2}
\end{figure}

The most important property of this conductance matrix is the very small value $10^{-4}$ of its upper-right element (dashed square in Fig.~\ref{fig1}d), which has a strong impact on the amplitude of the topological skin effect. The non-Hermitian topology of the conductance matrix is best revealed by the sum of probability densities (SPDs), calculated over all the eigenvectors of the matrix \cite{Ochkan2024}. Fig.~\ref{fig2}a) shows the exceptionally-large skin effect measured in zero magnetic field at base temperature (30mK), which exceeds by two orders of magnitude the amplitude of the skin effect measured with devices made from other quantum Hall systems \cite{Ochkan2024,Oezer2024}. At lowest temperatures, the residual bulk conductivity is fully suppressed, and the magnetic gap is large enough so that all resistances $R_i=V_i/I$ are stable over a wide field range (see Fig.~\ref{fig2}a, upper inset) and the skin effect remains unchanged, but in a narrow 20mT range at the magnetization reversal (see Fig.~\ref{fig2}a, lower inset). Two remarkable properties are the robust temperature dependence of the skin effect, which persists at 4.2K if a small magnetic field is applied (Fig.~\ref{fig2}b), and the rapid switching of the chirality at the coercive field, where the skin effect becomes extremely sensitive to a small change in the magnetic field (Fig.~\ref{fig2}c). For a 5-site chain, the resolution of this cryogenic magnetic-field sensor reaches the 10~nT range, given by the ratio of the maximum slope of about $2.10^{-1}$~mT$^{-1}$ to the accuracy of measurement of the SPD at the $10^{-6}$ level (grey areas in Fig.~\ref{fig2}). 
Increasing the current (improved signal-to-noise ratio) or the chain size (larger SPD) would be some appropriate strategies to futher improve the sensitivity below the 100~pT range.

{\bf Quantization of the topological invariants and temperature stability}

\begin{figure}[!t!h]
\includegraphics[width=\columnwidth]{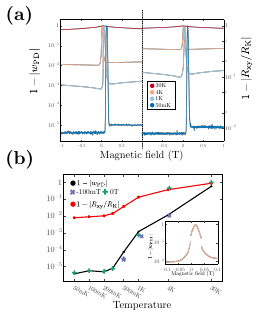}
\centering
\caption{
\textbf{Quantization of the non-Hermitian topological invariant and of the Hall resistance at various temperatures.} 
\textbf{(a)}  Magnetic-field dependence of $1 - \abs{w_{\text{PD}}}$ (left) and of $1 - \abs{R_{xy}/R_{\text{K}}}$ (right), where $R_{\text{K}} = h/e^2$, measured at different temperatures. 
\textbf{(b)}  Temperature dependence of $1 - \abs{w_{\text{PD}}}$ and $1 - \abs{R_{xy}/R_{\text{K}}}$, measured at a fixed magnetic field $B = -1.5$~T. The $w_{\text{PD}}$ is also shown at a lower field $B = -100$~mT and in zero field. Inset: Field dependence of $w_{\text{PD}}$ at 4.2K. A small applied field $\vert B\vert \geq 50$~mT is needed to maintain the magnetization saturated, with a good quantization of $w_{\text{PD}}$ only.
}
\label{fig3}
\end{figure}

The non-Hermitian skin effect is associated to a topological invariant, different from the Chern invariant, which can be calculated from the polar decomposition of the Hatano-Nelson model \cite{Claes2021,Ochkan2024}. As shown in Fig.~\ref{fig3}a, left), the invariant $w_{PD}$ related to the non-Hermitian topology is extremely well quantized, with a record value $4.10^{-6}$ for the deviation from perfect quantization, as measured for a 5-site device at $T=30$~mK. 
For comparison, the best quantization achieved with a similar quantum device implemented in AlGaAs 2DEG at large magnetic fields is worse by one order of magnitude, with a deviation $1-\vert w_{PD}\vert=8.10^{-5}$ for the $\nu=4$ quantum Hall plateau \cite{Ochkan2024}. The difference is even stronger when considering the SPD values for a 5-site chain, with an improvement for QAH devices related to better ohmic contacts. Still, even for QAH devices, the skin effect can be severely suppressed if the access resistance to the quantum device (setup wiring, contact quality) is not small enough \cite{Yi2025}.

Remarkably, the quantization of the non-Hermitian invariant is stronger than that of the Chern invariant, as measured in a standard lock-in amplifiers measurement setup without low-temperature filters (Fig.~\ref{fig3}a). Physically, this is intrinsic to the exponential localization at boundaries of an open quantum chain, with a strong decoupling between sites at opposite ends, which results in the vanishingly-small value $10^{-4}e^2/h$ of the upper-right corner term in the conductance matrix (see Fig.~\ref{fig1}d), whereas the degree of quantization of the Chern invariant (defined as the relative deviation of $R_{xy}$ from $R_K$), at the $10^{-3}$ level, is fixed by the residual backscattering between nearby successive sites. As a direct consequence, a striking property of the non-Hermitian device topology is its robustness with temperature. As Chern states start to coexist with thermally-activated bulk carriers, that is, above 300~mK, the quantization of both invariants gets worse (Fig.~\ref{fig3}b). However, contrary to the Chern invariant, the non-Hermitian topological invariant remains strongly quantized at 4.2K, yet with the need to apply a small magnetic field ($\vert B\vert \geq 50$~mT) in order to keep the magnetization saturated. This difference appears directly in the temperature dependence of some specific elements of the conductance matrix (see Fig.~\ref{fig:SI:G_matrices} for details). Importantly, the robustness of the topological skin effect with both imperfections and temperature sets the condition to use these multi-terminal devices as cryogenic sensors that can operate at liquid Helium temperatures. 
At even higher temperatures, the parallel bulk conductance is detrimental to the quantization of both invariants. A similar trend is found if a large magnetic field is applied, with a reduced temperature stability of all quantized properties related to the smaller magnetic gap, due the Zeeman effect (see Fig.~\ref{fig:SI:higherT}).


{\bf Influence of the device geometry and scaling.}


For large-size devices, there is no influence of the exact shape of the quantum device on the non-Hermitian topology of its conductance matrix, also if opposite-side edge states exist as well (hollow geometry of a ring-shaped device). The absence of crosstalk, with no impact on the edge-state backscattering length, is best evidenced by comparing the results obtained with a ring geometry to those measured with a uniform disk. Both the quantization of the non-Hermitian topological invariant and its temperature dependence are unchanged. This is not surprising since the backscattering length remains rather short in QAH devices \cite{Deng2022}, at the micron scale, so that the downscaling of such multi-terminal devices is possible. 

\begin{figure}[!t!h]
\includegraphics[width=\columnwidth]{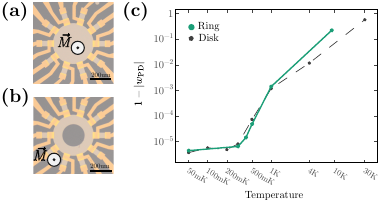}
\centering
\caption{
\textbf{Comparison of quantum disks and rings} 
 Multi-terminal quantum disk \textbf{(a)} and quantum ring \textbf{(b)}, and comparison of the temperature dependence of their non-Hermitian topological invariant \textbf{(c)}. 
}
\label{fig4}
\end{figure}

\section{\label{Discussion}Discussion}
With their outstanding performance, both in terms of quantization and robustness in temperature or with imperfections, QAH-based quantum devices with non-Hermitian topology are a new generation of topological devices with functionalities as cryogenic sensors. These devices can operate in zero or small magnetic fields, in an industry-compatible environment, with no need of advanced measurements techniques nor of ultra-low temperature cryogenics. Due to the exponential sensitivity of the skin effect to boundary conditions, such quantum devices can be used as high-impedance meters, and the possibility to switch at will the chirality at small magnetic field gives some additional functionality as magnetic-field sensors. New opportunities can be considered for fundamental research as well, for instance with the local control of magnetism in magnetic topological insulators, in order to tailor the conductance matrix and engineer other Hamiltonians, beyond the Hatano-Nelson model.

\section{\label{Methods}Methods}


{\bf Quantum transport measurements}

Magnetotransport measurements were performed with ac lock-in amplifiers, using a 20nA polarization current at $f=$7Hz. The samples were mount onto the cold finger of an Oxford Instruments MX400 $^3$He-$^4$He dilution refrigerator (28mK base temperature), fitted into a 15T superconducting-coil magnet, with a 280~$\Omega$ wiring resistance in series for the current injection.

{\bf Calculation of the non-Hermitian topological invariant}

The non-Hermitian skin effect has a topological origin \cite{Okuma2020}, meaning that it is the consequence of a nonzero bulk topological invariant.
While the latter is usually expressed in terms of a winding number of the Hamiltonian as a function of momentum, here, in view of the finite size of the conductance matrices, we use a real-space formulation of the winding number.
Following Refs.~\cite{Herviou2019,Claes2021,Bergholtz2021}, we calculate the polar decomposition the OBC conductance matrix as
\begin{equation}
    G_\text{OBC} - \text{Tr}(G_\text{OBC})= QP,
\end{equation}
where $Q$ is a unitary matrix, $P$ is positive, and Tr denotes the trace.
The real-space winding number is then expressed as
\begin{equation}
    w_\text{PD} = {\cal T}(Q^\dag [Q, X]),
\end{equation}
where $X=\text{diag}(1, 2, \ldots, L)$ labels the active arms of the device (sites/terminals used as current or voltage probes), and ${\cal T}$ is the ``trace per unit volume.''
In practice, this corresponds to a sum over diagonal elements corresponding to sites (or terminals) away from both ends of the effective Hatano-Nelson chain.





\paragraph{\bf Acknowledgements} 
This work was supported by the European Union’s H2020 FET Proactive project TOCHA
(No. 824140), as well as by the CNRS International Research Project``CITRON".

\let\oldaddcontentsline\addcontentsline
\renewcommand{\addcontentsline}[3]{}
\bibliography{NHQAH.bib}
\let\addcontentsline\oldaddcontentsline

\clearpage


\beginsupplement
\section*{Supplementary Information} 

\subsection{Temperature dependence of the resistance and conductance matrices (disk geometry)} \label{sec:matrices_for_Ts}

\begin{figure}[!t!h]
\centering
\includegraphics[width=5.6cm]{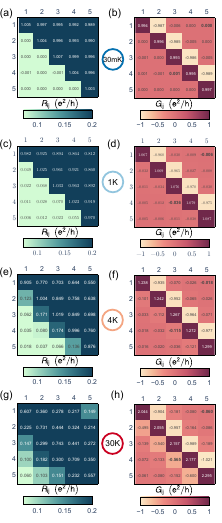}
\caption{
\textbf{Resistance and conductance matrices} measured at $T= 30$~mK (a,b), $T= 1$~K (c,d), $T= 4$~K (e,f), and $T= 30$~K (g,h), from which the $w_{\text{PD}}$ and $R_{\text{xy}}$ parameters are inferred, as shown in Fig.~\ref{fig3}. $B = -1.5$~T.
}
\label{fig:SI:G_matrices}
\end{figure}

Fig.~\ref{fig:SI:G_matrices} shows the evolution with temperature of the resistance and conductance matrices, as measured for the disk-shaped device, from which the SPD and the invariants are calculated, as shown in Fig.~\ref{fig2} and Fig.~\ref{fig3}. 

At lowest temperature, in the QAH regime, the conductance matrix realizes nearly ideally the Hatano-Nelson Hamiltonian situation, with zero elements apart from the diagonal and the first upper-diagonal terms (quantization of the conductance for Chern states). Upon thermal activation of the parallel bulk conductance, some finite-value elements develop in the conductance matrix. Most important to consider are the first lower-diagonal terms, related to the quantization of the Chern invariant, and the upper-right corner term, related to the non-Hermitian topological skin effect and to the quantization of the $w_{\text{PD}}$ invariant (see bold-font elements in Fig.~\ref{fig:SI:G_matrices}b,d,f,h). The non-zero amplitude of the upper-right corner term 
at higher temperature is smaller than the change of other terms, so that the deviation from perfect quantization is much stronger for the Chern invariant than for the device non-Hermitian topological invariant, a finding that is best summarized in Fig.~\ref{fig3}b).

\subsection{Magnetoresistance at higher temperatures (disk geometry)} \label{sec:higherT}

\begin{figure}[!t!h]
\centering
\includegraphics[width=6cm]{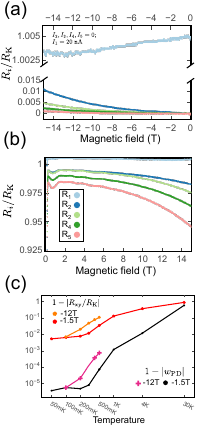}
\caption{
\textbf{High-field behavior.} 
Resistances measured at $T = 200$~mK in the configuration shown in Fig.~\ref{fig1}a) under a field sweep from -15~T (a) and up to 15~T, after magnetization reversal at small positive fields (b). (c), Temperature dependence of $1 - \abs{w_{\text{PD}}}$ and $1 - \abs{R_{xy}/R_{\text{K}}}$, measured at $B=$-1.5~T or -12~T. 
}
\label{fig:SI:higherT}
\end{figure}

While the residual bulk conductivity is fully suppressed at lowest temperatures, with all resistances stable over a wide field range in the QAH regime (upper inset in Fig.~\ref{fig2}a), the situation is different at higher temperatures, that is, above $300$~mK. 
Due to the reduction of the small magnetic gap by the Zeeman effect, this onset of parallel bulk conductivity is best seen at high fields, leading to deviations from perfect quantization and a finite longitudinal resistance (see Fig.~\ref{fig:SI:higherT}a,b). As expected, the non-Hermitian topological invariant has a reduced onset temperature as well (see Fig.~\ref{fig:SI:higherT}c). Still, similar to the results found for the temperature dependence, the degree of quantization of this invariant remains very large in high magnetic fields, whereas the deviation from the quantized Chern invariant is large.

\subsection{Influence of the grounding scheme and contact spacing (ring geometry)} 
\label{sec:grounding_sites}

\begin{figure}[!h!t]
\includegraphics[width=\columnwidth]{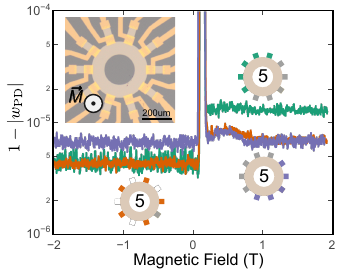}
\centering
\caption{
\textbf{Non-Hermitian topology for a QAH ring device.} 
Influence of the contact configuration on the ultimate performance of the ring-shaped QAH devices for the 5-site configuration, as assessed by the accuracy of the invariant quantization, measured at $T = 30$~mK. Neither the exact position of the contacts (colored squares) nor the number of contacts (grey squares) used to ground the device alter the degree of quantization of the device non-Hermitian topological invariant. Empty-square contacts are floating; Inset: Optical microscopy picture of the QAH ring device. 
}
\label{fig:SI:NHSE_and_ground}
\end{figure}

Finite size effects in a multi-terminal non-Hermitian topological device are often hidden by finite contact resistances and access resistances of the experimental setup, so that the grounding scheme plays a similarly important role for the degree of quantization of their invariant. 
This is exemplified in Fig.~\ref{fig:SI:NHSE_and_ground}, considering different contact configurations for the 5-site quantum chain. The performance of the topological device, in terms of quantization of the invariant, varies only slightly with the exact position of contacts or with their relative spacing. It is also rather independent from the exact number and position of contacts used to ground the device. Although small variations in the contact resistance from contact to contact can give a small asymmetry in the field dependence, that is, for opposite chirality, the overall device performance always remains at the state of the art.

\end{document}